\documentstyle[pre,aps,tighten,multicol,epsfig]{revtex}

\begin{document}
\draft

\psfull

\title{Non glassy ground-state in a long-range antiferromagnetic frustrated
model in the hypercubic cell}
\author{Leonardo Franco}

\address{University of Oxford, Dept. of Experimental Psychology, South Parks
Road, Oxford OX1 3UD, United Kingdom}

\author{Sergio  A. Cannas\cite{auth2}}

\address{Facultad de Matem\'atica, Astronom\'\i a y F\'\i sica,  Universidad
Nacional de C\'ordoba, Ciudad Universitaria, 5000   C\'ordoba, Argentina}

\date{\today}

\maketitle

\begin{abstract}

We analize the statistical mechanics of a long-range antiferromagnetic model
defined on a $D$-dimensional hypercube, both at zero and finite temperatures.
The associated Hamiltonian is derived from a recently proposed  complexity
measure of Boolean
functions, in the context of neural networks learning processes.
We show that, depending of the value of $D$, the system either presents a low
temperature antiferromagnetic stable phase or the global  antiferromagnetic
order
disappears at any temperature. In the last case the ground state  is an
infinitely degenerated non-glassy one, composed by two equal size anti-aligned
antiferromagnetic domains.
We also present some results for the ferromagnetic version of the model.

\end{abstract}



\section{Introduction}

The success in understanding the collective behavior of conventional magnets
lead
in the past decade to an increased interest in systems that exhibit a novel type
of ordering as a consequence of ``frustration''. In
particular, geometrically frustrated systems (having as a basic building block
triangles of antiferromagnetic bonds) have been the object of extensive
theoretical
 studies (see \cite{Ogawa} and references therein). It was found that
frustration gives rise to
spectacular and often unexpected behaviors at low temperatures due to the high
 degeneracy of the ground state. Recently, more and more experimental
realizations
of such geometrically frustrated systems have
been achieved. 2D and 3D materials have been assembled from triangles and
tetrahedra such that neighbouring geometrical units share either a common
edge or a common corner \cite{Girtu}. Such new materials have stirred a
new debate, whether frustration alone is strong enough to destroy magnetic
long range order.
Spin glass behavior is usually associated to the presence of  both disorder and
frustration.
 It has been argued that frustration alone can lead to large ground-states
degeneracies, accompanied by extensive ground-state entropies, but
cannot produce a sufficiently ``rough'' free energy landscape necessary for
 the development of a thermodynamical "glassy" state\cite{Chandra,Rammal}.

In contrast with the previously mentioned systems, non-geometrical frustration
may arise as an
 effect of antiferromagnetic long-ranged interactions (e.g., dipolar, Coulomb,
etc.). It is well known
that the competition between short-range ferromagnetic exchange interactions and
long-range antiferromagnetic
 ones in hypercubic Bravais lattices give rise to low temperature long range
order associated with
lamellar structures\cite{DeBell,Grousson}. Although those systems presents a
very rich behaviour
concerning both its equilibrium\cite{DeBell}-\cite{Grousson} and
non-equilibrium\cite{Toloza,Grousson2}
 properties, in all these cases the ground state presents a finite degeneracy.
Even in the case of a 3D system with
 Coulomb interactions, where  some evidence of glassy behaviour has been
obtained\cite{Grousson2},
 it is of purely dynamical origin, the ground state still having a finite
degeneracy.

In this work we analize the zero and low temperature equilibrium properties of a
long range 
antiferromagnetic spin model defined on a $D$-dimensional hypercubic cell. The
associated
 Hamiltonian is  related to a recently proposed complexity measure of Boolean
functions
 in the context of neural networks learning processes and it is described in
section \ref{model}. Although the
original motivation for
 studying the statistical mechanics of this model is related to the the field of
Boolean functions
complexity and neural networks \cite{Franco}, we found that it presents some
very peculiar
properties that provide new insights about the relationship between frustration
and order-disorder low temperature properties. 
Spin models defined on hypercubic cells have been repeatedly used in the past to study both
equilibrium and
dynamical properties of spin glasses\cite{Parisi}-\cite{Stariolo2}.  In that context, the main interest in hypercubic cell models is to analyze how their properties change as the dimension (and therefore the connectivity of the lattice) increases, thus resembling fully connected models like the Sherrington-Kirkpatrick one\cite{Sherrington}, whose general properties are rather well understood\cite{Parisi2,Cugliandolo}. One of the main questions today concerning spin glasses is whether the physical scenario that emerges in long-range fully connected models holds in systems with short range interactions or not. One step further towards the answer of that question is to consider systems with long range interactions that decay with the distance between spins. In this work we analize the effects of frustrating long range antiferromagnetic interactions in an hypercuibc cell without disorder. The  interactions in our model  decay with the Hamming distance between spins, a natural metric in hypercubic cells, instead of the Euclidean distance. Notice that the Hamming distance in the hypercubic cell is equivalent to the chemical distance (i.e., the number of steps required to go from one node to another along the shortest path). Hence, the present study may also provide some insigths about the possible effects of frustrating interactions we  expect in models defined on lattices with complex network topologies, such as small world networks (see Ref.\cite{Strogatz} and references therein).

We show that, for some particular values of the dimension $D$ the frustrating
long range interaction gives rise to
a ground state that is neither disordered (i.e., non-glassy) nor completely
ordered but, it is composed of two equal size
anti-aligned antiferromagnetic domains. The ground state is thus partially
ordered. Moreover, it has an infinite
degeneracy in the thermodynamic limit (although this degeneracy grows
logarithmically with the number of spins,
and hence its entropy per spin goes to zero).

In section \ref{ground} we analyze the frustrated ground state of the system for
the particular set of values of $D$ such that it exhibit global frustration.
In section \ref{T} we analize the finite temperature behaviour of the model
using Monte
Carlo simulations.
 In section \ref{ferro} we present Monte Carlo results for the ferromagnetic
version of the model,
which are of interest in the general context of long range interacting systems.
Finally, we present some general conclusions  in section \ref{conclu}.

\section{The model}

\label{model}

In Ref.\cite{Franco} a measure for the complexity of Boolean functions of $D$
Boolean variables
was introduced in the context of feed-forward neural networks learning
processes.  The $2^D$ possible
 inputs of a Boolean function of this type are located at the corners of a
$D$-dimensional  hypercubic cell.
 If we associate a spin variable $ S_i = \pm 1$ to each corner $i$ of the cell
every  possible Boolean
function corresponds to a particular spin configuration of this system. The
complexity measure is based on the
 concept that complex Boolean functions assign different outputs to similar
inputs, where the similarity is
 measured by the Hamming distance between them. Thus, in the magnetic analogy,
configurations where
neighbour spins tend to be anti-aligned correspond to functions of increasing
complexity. The complexity
measure can be then mapped into minus the long range antiferromagnetic
Hamiltonian\cite{Franco}

\begin{equation}
H =  \sum_{i,j} J(r_{ij}) S_i S_j
\label{Hamiltoniano}
\end{equation}

\noindent where $r_{ij}$ is the Hamming distance between spins $i$ and $j$ in
the hypercubic cell and

\begin{equation}
 J(r)  = \left\{ \begin{array}{ll}
   \frac{1}{M(r)} & \mbox{if $ r \le D/2$} \\
   0 & \mbox{otherwise}
 \end{array}
 \right.
\label{J(r)}
 \end{equation}

\noindent $M(r)$ being the number of nearest neighbors at distance $r$ of a
given spin and  equals
the binomial coefficient

\begin{equation}
M(r)= C(D,r)= \frac{D!}{(D-r)! \; r!} .
\end{equation}

\noindent The distribution of Boolean functions with high complexity is then
associated with the low energy spectrum
of Hamiltonian (\ref{Hamiltoniano}) which can be characterized by the  low
temperature statistical mechanics behaviour of the associated spin model.
The Hamiltonian (\ref{Hamiltoniano}) can be rewritten as a sum of terms,
each of them taking into account
the interactions between pair of spins at a different Hamming distance:

\begin{equation}
H = \sum_{r=1}^{r= D/2} J(r) \sum_{ [i,j / Hamming(i,j)=r ] }  S_i S_j ,
\label{eq_hamilt}
\end{equation}

Notice that the number of interaction terms in  Hamiltonian (\ref{eq_hamilt})
depends on $D$; as $D$ increases
 a new term is included every time $D$ takes an even value. For instance, in the
cases $D=2$ and $3$  only nearest
neighbor spins interact; in the cases $D=4$ and $5$ an interaction between next
nearest neighbor spins is added, etc..
 Hence, the antiferromagnetic nature of the interactions introduce local
frustration for $D \geq 4$, in the sense that
 not all the couplings inside groups of spins closely located can be
simultaneously minimized. However, this does not
necessarily generate global frustration, in the sense of a non-ordered ground
state with high degeneracy, due to  the
particular form Eq.(\ref{J(r)}) of the interaction coupling $J(r)$. Global
frustration appears only when the number
of interaction terms included in the Hamiltonian (\ref{eq_hamilt}) is even,
that is, when the dimension $D$ is a
multiple of four $D = 4 l $, $l=1,2,3,\ldots$. In all the other cases the ground
state is the antiferromagnetic one
 (i.e., the N\`eel state with every pair of nearest neighbor spins
anti-aligned), with degeneracy $2$ (corresponding to
 the inversion of all the spins). To see this let us calculate the energy of the
N\`eel state. Every term in Eq.(\ref{eq_hamilt})
corresponding to a fixed Hamming distance $r$ equals in this case $(-1)^r N\;
M(r)$, $N=2^D$ being the total number of
 spins (notice that the interaction between every pair of spin in Hamiltonian
(\ref{Hamiltoniano}) is counted twice for
simplicity; this corresponds to a renormalization of the energy units by a
factor two, and does not affect the results).
Hence, the energy of the N\`eel state is given by

\begin{equation}
E_N = N \; \sum_{r=1}^{r= D/2} (-1)^r
\end{equation}

\noindent which equals zero when $D/2$ is even and $E_N/N=-1$ when $D/2$ is odd.
Moreover, in the last case is easy to see that a single spin flip increases the
energy in a quantity $\Delta E=2$. Therefore, the N\`eel state is stable against
small perturbations. Monte Carlo simulations presented in section \ref{T}
provide further evidence that the ground state is the N\'eel one. Moreover, the
finite temperature simulations show that  there is a finite critical temperature
below which the system presents long range antiferromagnetic order when $D/2$ is
odd.

In the next section we analyze the frustrated ground state for the case  $D = 4
l $, $l=1,2,3,\ldots$.

\section{The frustrated ground state}

\label{ground}

\begin{figure}
\begin{center}
\psfig{figure=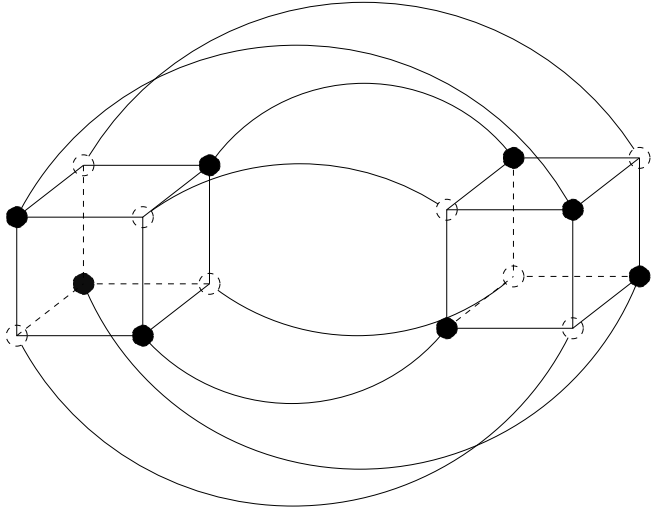,width=10.00cm}
\end{center}
\caption{Ground state of the antiferromagnetic Hamiltonian for $D=4$ . Spin
states are represented by empty or filled circles and the bond between spins
indicate that they are first nearest neighbors.}
\label{fig_cubo_D4}
\vspace{0cm}
\end{figure}

From exhaustive enumeration of the energy values for the case $D=4$, where the
Hamiltonian contains two terms, it was
found that the ground state of the  system consist of two equal size
antiferromagnetic domains with an energy value equal
to $-0.5$. Monte Carlo simulations in higher dimensions ($D=4,8,12,16$) have
also shown that the energy value of the ground
state of the  system is equal to $-0.5$ and has the same structure found for the
case for $D=4$. (Long relaxation times and
very low cooling were used  in different simulations, all of them confirming the
structure and energy value of the ground state).
 Energy calculations for states composed by $4$ and $8$ anti-aligned
antiferromagnetic domains($D=8,16$) also yield
values higher than $-0.5$. Moreover, as we will demonstrate next, the energy per
spin of the state composed by two antiferromagnetic domains equals $-0.5$ for
arbitrary
values of $D$. Since the antiferromagnetic state has always zero energy, all
these result suggests that the two antiferromagnetic
domains  structure has the minimal energy for arbitrary dimension $D=4l$,
$l=1,2,3,\ldots$.
The total magnetization and staggered magnetization is zero for the ground
state.  In figure \ref{fig_cubo_D4}
a representation of the ground state for $D=4$ is depicted, where
the state of the spins is indicated by filled or empty circles.
 The antiferromagnetic domains correspond to two hypercubes of dimension $D-1$
each
 (see Fig. \ref{fig_cubo_D4}).  Each spin in this state has $D-1$ nearest
neighbor spins anti-aligned,
which belong to the same domain, and one nearest neighbor spin aligned, which
belong to the other
domain. Once we fixed two nearest neighbor spins to be aligned, the rest of the
structure is completely
determined. Since every spin has $D$ nearest neighbors, the degeneracy of the
state is $2\, D=2 \; {\rm log}_2 N$,
where the factor $2$ results from the inversion operation of all the spins.

We now demonstrate that the energy per spin $E_d$ of the state composed of two
antiferromagnetic domains  equals  $-0.5$ for any dimension $D= 4l$,
$l=1,2,3,\ldots$. $E_d$ can
be written as

\begin{equation}
E_d = \sum_{n=1}^{n=D/2} E_d(n)
\label{Ed}
\end{equation}

\noindent $E_d(n)$ being the interaction energy between an arbitrary chosen
reference spin and all its neighbors located at a Hamming distance $n$.

We denote the two antiferromagnetic domains as regions  A and B.  Consider now
an arbitrary spin $S_1$ that belongs to the domain A. Spin $S_1$ has  $D-1$
anti-aligned  nearest neighbors (NN) spins in the domain A and only one first
nearest neighbor spin  $S_2$ in the domain B. Spin $S_2$ has  the same
orientation than the spin $S_1$. As the total number of  NN spins in dimension
$D$ is exactly $D$, the first term of Eq.(\ref{Ed}) equals

\begin{equation}
E_d(1) = -  \frac{(D-1)}{D} + \frac{1}{D}
\end{equation}

\noindent which can be expressed as

\begin{equation}
E_d(1) =  2 \frac{C(D,0)}{C(D,1)} - 1
\end{equation}

Let us now consider the next nearest neighbors (NNN) term $n=2$ in
Eq.(\ref{Ed}). Note that the NNN spins of $S_1$ in the domain B (anti-aligned to
$S_1$) are the $D-1$ spins NN of $S_2$ in B; the remaining spins NNN to $S_1$ in
A have the same orientation of it. Since the total number of NNN spins is
$C(D,2)$ we have that

\begin{eqnarray}
E_d(2) &=& -\frac{(D-1)}{C(D,2)} + \frac{C(D,2) - (D-1)}{C(D,2)} \nonumber\\
          &=&  - 2 \frac{C(D,1)-C(D,0)}{C(D,2)} + 1
\end{eqnarray}

\noindent Following the same analysis the $3rd$ term of Eq.(\ref{Ed}) equals

\begin{equation}
E_d(3) =  2 \frac{C(D,2)-C(D,1)+C(D,0)}{C(D,3)} -1
\end{equation}

\noindent and the $n-th$ order term can be written as

\begin{equation}
E_d(n) =  2 \frac{\sum_{k=0}^{n-1} (-1)^k C(D,k)}{C(D,n)} + (-1)^n.
\label{Edn}
\end{equation}

\noindent Equation (\ref{Edn}) can be rewritten as

\begin{equation}
E_d(n) =   (-1)^{n+1} \left( 2\, a(n) -1 \right)
\label{Edn2}
\end{equation}

\noindent where

\begin{equation}
a(n) \equiv   (-1)^{n+1} \frac{\sum_{k=0}^{n-1} (-1)^k C(D,k)}{C(D,n)}.
\label{a}
\end{equation}

\noindent It is s easy to verify that $a(n)$ satisfy the recurrence equation

\begin{equation}
a(n+1) =  \frac{C(D,n) (1 - a(n))}{C(D,n+1)}  = \frac{n+1}{D-n} \left( 2\, a(n)
-1 \right)
\end{equation}

\noindent from which it can be proved by induction that $a(n)=n/D$. Replacing
into Eqs.(\ref{Ed}) and
(\ref{Edn2}) we finally obtain

\begin{eqnarray}
E_d  &=&  \sum_{n=1}^{n=D/2}(-1)^{n+1}\left( \frac{2n}{D}-1 \right) \nonumber\\
\mbox{}  &=&  -\, \frac{2}{D} \sum_{n=1}^{n=D/2}(-1)^{n} \; n   = -\,
\frac{1}{2}
\end{eqnarray}

\noindent where in the last step we have used that

\begin{equation}
 \sum_{n=1}^{n=D/2} (-1)^n n=\frac{D}{4}
\end{equation}

\section{Finite temperature equilibrium properties}

\label{T}

\begin{figure}
\begin{center}
\psfig{figure=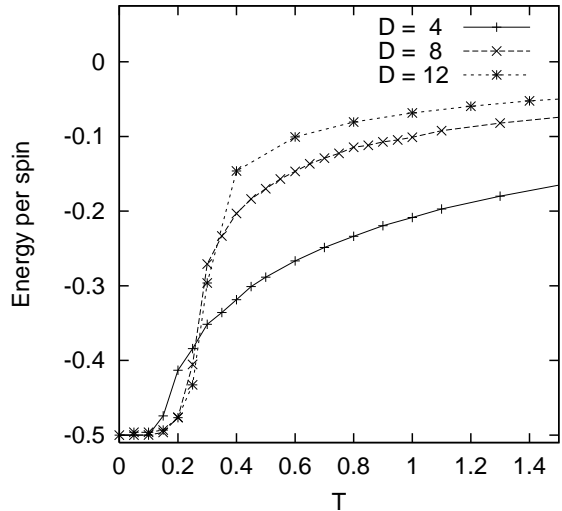, width=8.00cm}
\end{center}
\caption{Monte Carlo calculation of the mean energy per spin vs  temperature for
an antiferromagnetic Hamiltonian with $D=4,8,12$.
\label{fig_AFe_fr}}
\vspace{0cm}
\end{figure}

We now present some numerical simulation results at finite temperature.
 Monte Carlo simulations were performed using heat bath dynamics.
We first analyze the frustrated case $D=4l$, $l=1,2,\ldots$. In Fig.
\ref{fig_AFe_fr}
we show the results for the internal energy per spin for system sizes
$D=4,8,12$.
The different curves were obtained by starting at high temperature with random
or
 ferromagnetic initial conditions (there were no difference between both cases
in the final results)
and slowly cooling, that is, the initial configuration at every subsequent
temperature was taken as
 the last configuration of the previous temperature. At every temperature we
left the system
 thermalize during a varying transient period, which was  higher at low
temperatures.

For low values of $D$ we see that the energy converges to the value $-0.5$.
Direct inspection of
the final configurations obtained at $T=0$ shows always the conjectured state
composed by two
antiferromagnetic domains. Notice from Fig.\ref{fig_AFe_fr} that there seems to
be a change of behaviour around $T\sim 0.2$.
 An analysis of the energy distribution at low temperatures showed that this
effect is related to a gap
between the ground state and the first excited states that decreases as $D$
increases and hence, it is a finite size effect.
Both the magnetization and the staggered magnetization are always ${\cal
O}(N^{-1/2})$ at any
 temperature and  also the specific heat does not show any type of anomaly. So
the system does not (globally) order at any temperature.

We next performed Monte Carlo simulations for the antiferromagnetic system when
 $D/2$ is odd.

\begin{figure}
\begin{center}
\psfig{figure=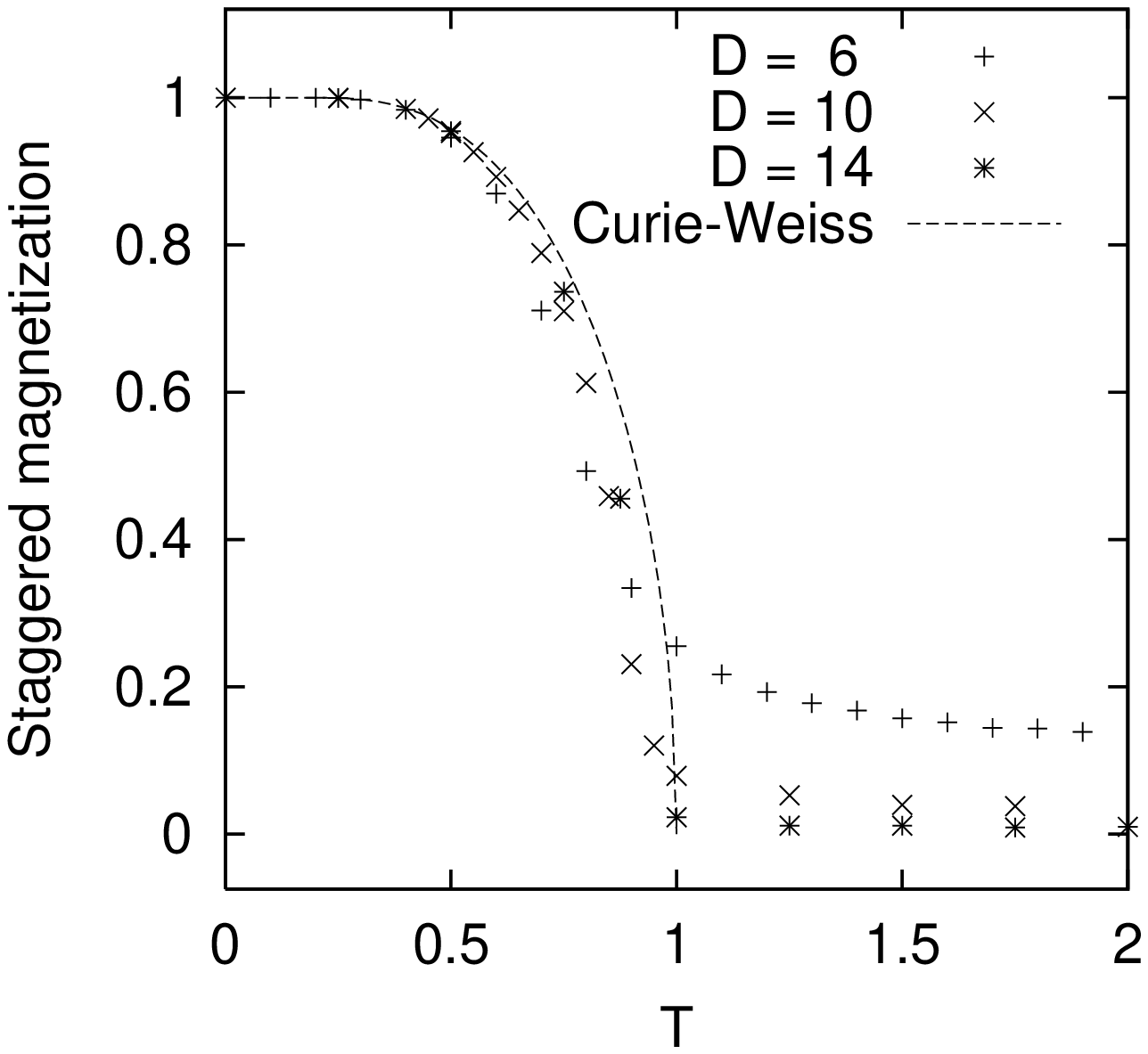, width=8.00cm}
\psfig{figure=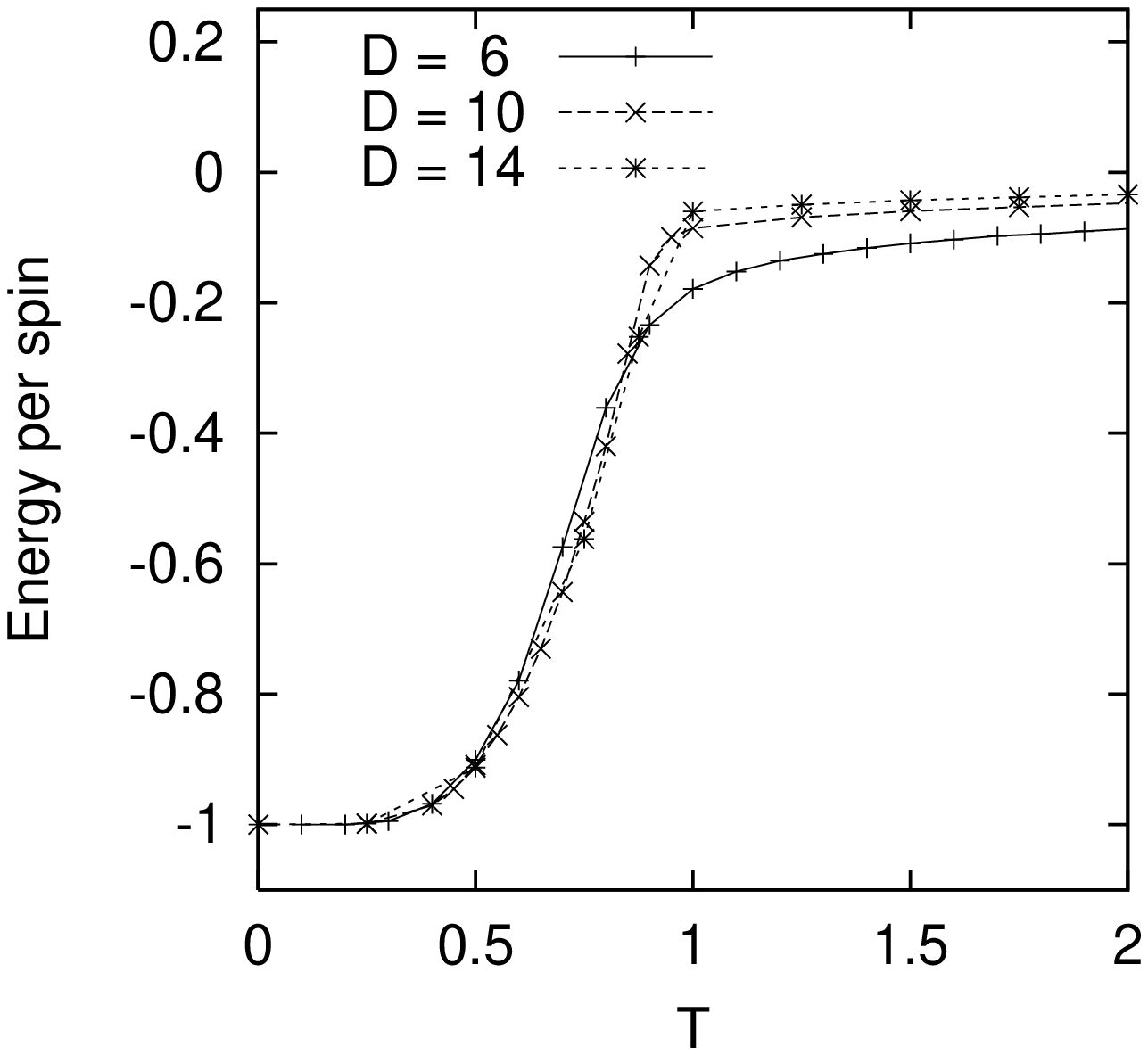, width=8.00cm}
\psfig{figure=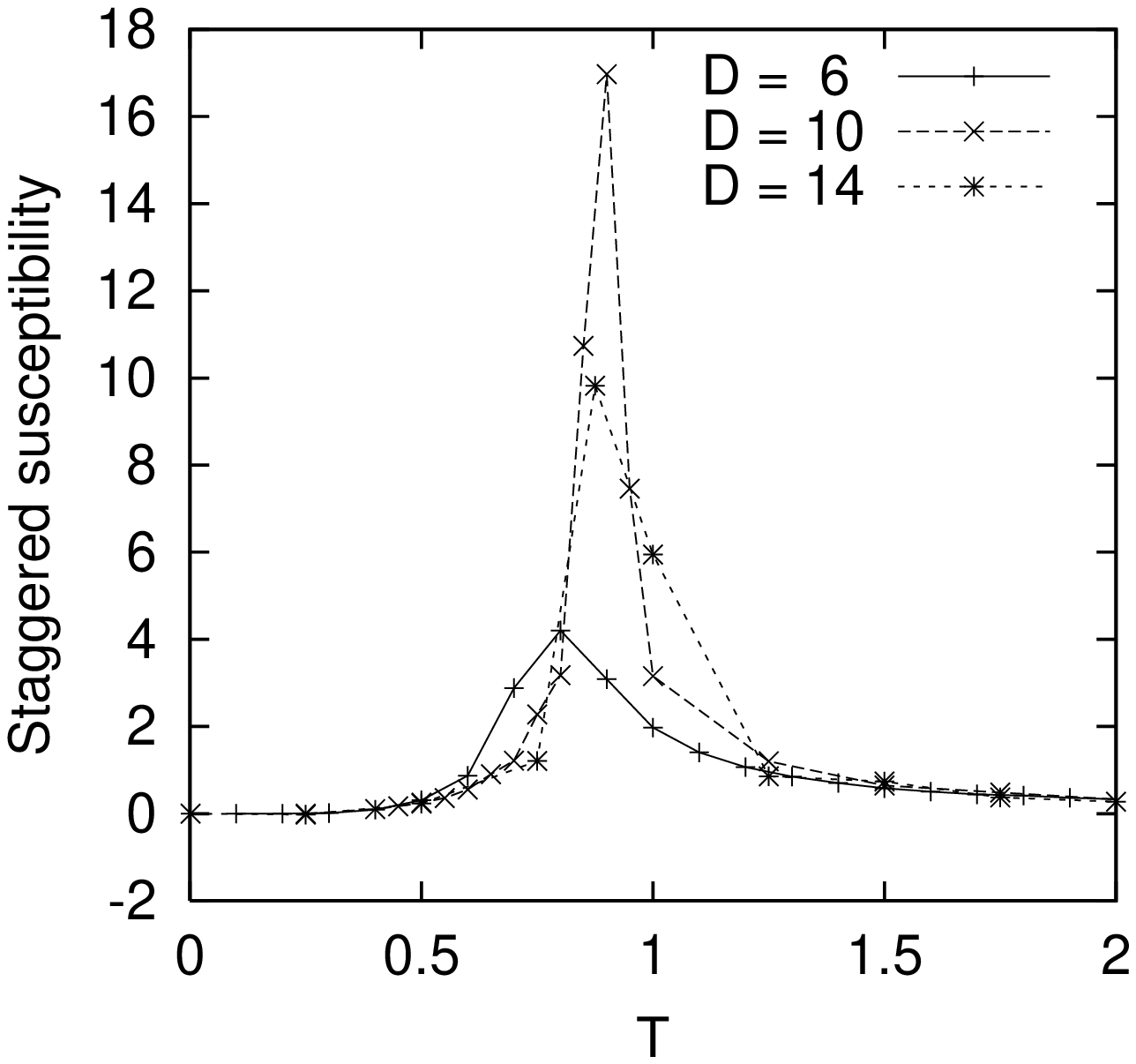, width=8.00cm}
\end{center}
\caption{Monte Carlo calculation of the mean staggered magnetization per spin,
energy per spin and staggered susceptibility vs
temperature  for an antiferromagnetic Hamiltonian with $D=6,10,14$.
 \label{fig_AFe_nfr}
}
\vspace{0cm}
\end{figure}

In figure \ref{fig_AFe_nfr} we show  the absolute value of the staggered
magnetization,  the energy per spin and the staggered susceptibility vs
temperature  for $D=6,10,14$. These results not only confirms that the ground
state is antiferromagnetic, but also show that the system undergoes a second
order phase transition at a finite critical temperature $T_c \approx 1$. The
comparison of the magnetization curves with the solution of the Curie-Weiss
equation $m=tanh(\beta m)$ suggests that mean field behaviour is exact in this
case at any temperature.

\section{The ferromagnetic model}

\label{ferro}

Finally we consider the ferromagnetic version of the model, that is, the model
defined by the Hamiltonian (\ref{Hamiltoniano}) with $J(r)= -\frac{1}{N(r)}$.
This system is non-extensive, in the sense that the quantity

\begin{equation}
\phi = -\sum_{j\ne i} J(r_{ij})=D/2 =  \frac{\log_2 (N)}{2}
\end{equation}

\noindent diverges for $N\rightarrow\infty$, and therefore the thermodynamic
limit is not defined\cite{Thompson,Ainzenman}.
 In this situation the magnetization per spin $m(T)$ and the internal energy per
spin $u(T)$ are expected to scale
as\cite{Tsallis,Cannas} 

\begin{eqnarray}
m(T)&\sim& m'(T/\phi) \nonumber \\
u(T)&\sim& \phi u'(T/\phi) \label{scaling} 
\end{eqnarray}

\noindent when $N\gg 1$, $m'(x)$ and $u'(x)$ being
scaling functions independent of $N$. In the case of $d$-dimensional hypercubic {\it lattices} with interactions that decay with the distance $r$ between spins as $1/r^\alpha$ the quantity $\phi$ is proportional to\cite{Cannas}

\begin{equation}
N^* = \frac{1}{1-\alpha/d} \left( N^{1-\alpha/d} -1 \right)
\label{N*}
\end{equation}

\noindent This scaling factor generalizes the traditional scaling $J\rightarrow J/N$ of the couplings in the Curie-Weiss model, i.e., a fully connected Ising spin system with distance-independent interactions, which corresponds to the $\alpha=0$ case. For $\alpha>d$ the scaling factor (\ref{N*}) becomes independent of $N$ in the thermodynamic limit and the system is extensive, while for $\alpha\rightarrow d$ (this corresponds, for instance, to the case of three dimensional systems wtih dipolar interactions that decay as $1/r^3$ for $r \gg1$)  we have that $N^*\rightarrow \ln{N}$. Thus, the behavior of the present hypercell model is expected to be analog to that of the borderline case between short and long range interactions $\alpha=d$  in hypercubic Bravais lattices.

\begin{figure}
\begin{center}
\psfig{figure=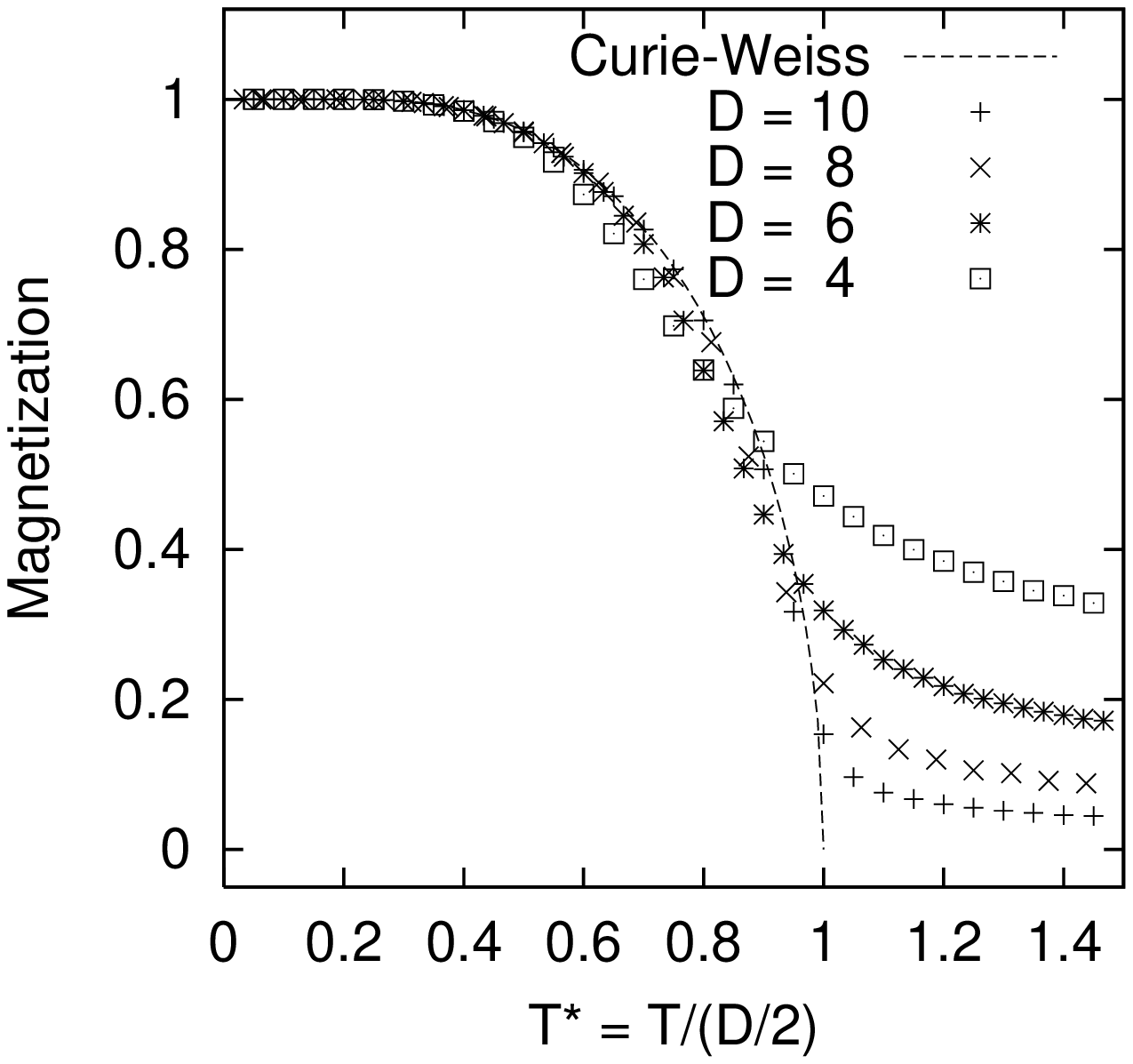, width=8.00cm}
\psfig{figure=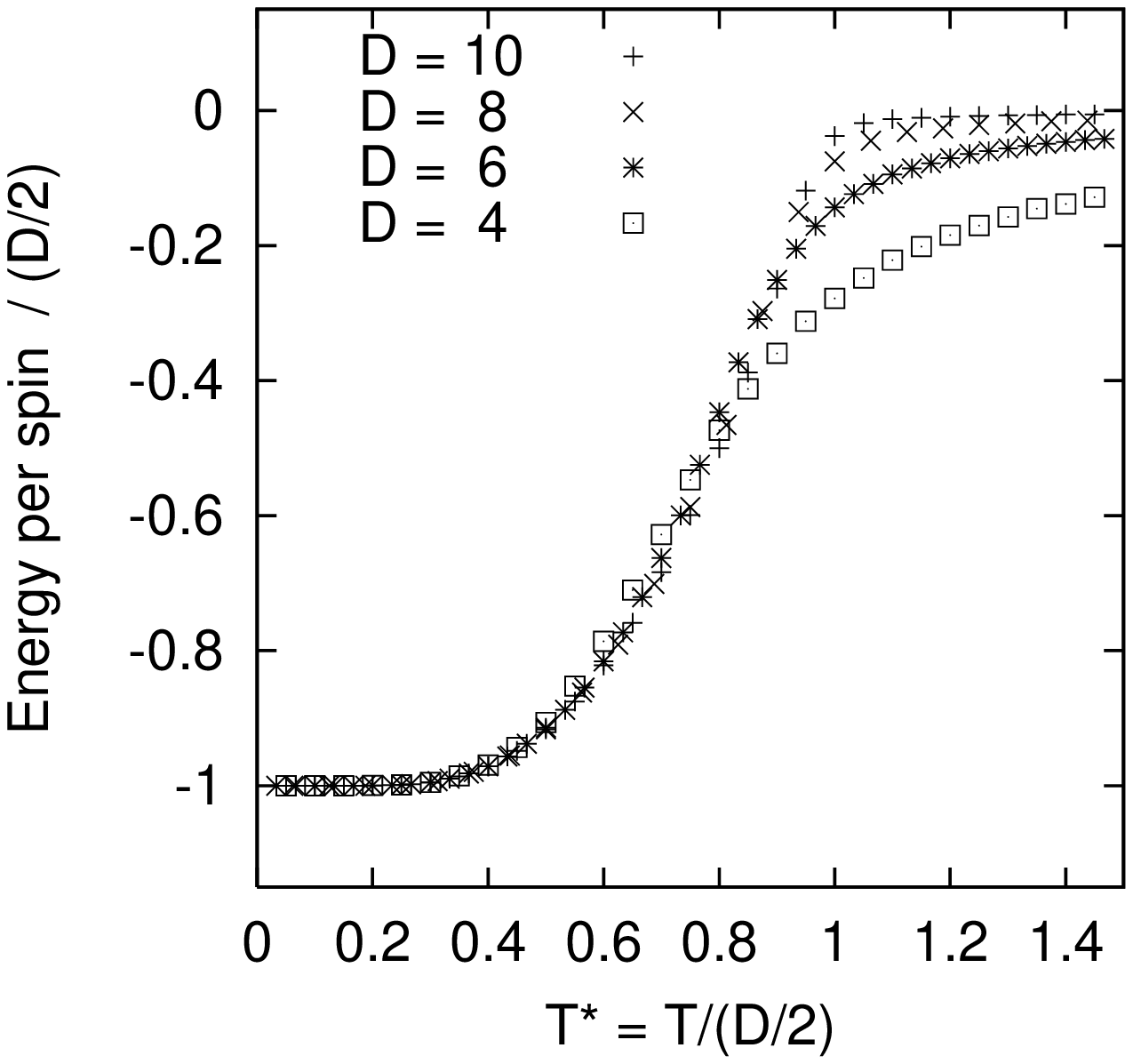, width=8.00cm}
\end{center}
\caption{Monte Carlo calculation of the mean magnetization per spin and energy
per spin vs
rescaled temperature $T^*$   for a ferromagnetic Hamiltonian with $D=4,6,8,12$.
 \label{fig_Fe}
}
\vspace{0cm}
\end{figure}

\begin{figure}
\begin{center}
\psfig{figure=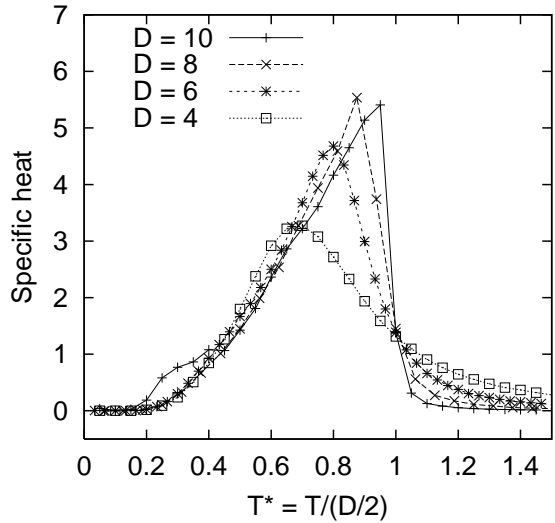, width=8.00cm}
\end{center}
\caption{Monte Carlo calculation of the specific heat vs  rescaled temperature
$T^*$ for $D=4, 6,8,10$.
\label{fig5}}
\vspace{0cm}
\end{figure}

The data collapse of the  rescaled
quantities for different values of $D$ shown
in Fig.(\ref{fig_Fe}) verify the scaling (\ref{scaling}). Moreover, the comparison of the
magnetization curves with the solution of the
Curie-Weiss equation $m=tanh(m/T^*)$ (where $T^*\equiv T/\phi$) shows that  mean
field behavior is asymptotically exact
for  a system of this type when the thermodynamic variables are properly
rescaled, as was previously conjectured for general
 systems with long-range ferromagnetic interactions\cite{Cannas2}.
Also the specific heat calculation shown in Fig.(\ref{fig5}) displays a
convergence to a discontinuous curve at the rescaled
critical temperature when $N\rightarrow\infty$, consistently with the mean field
behaviour.

\section{Discussion}

\label{conclu}

We analyzed the statistical mechanics of a long-range model defined on a
$D$-dimensional hypercube, which is related to the complexity of Boolean
functions. Besides its interest in Boolean functions complexity and neural
networks fields, the model exhibits several features that contribute to
the understanding of general long-range systems. Thus, the statistical
mechanics of the model is of interest on its own.

In the antiferromagnetic (AF) version of the model, the particular form of the
coupling terms in Eq.(\ref{J(r)}) actually define two different models, 
according to the value of $D$. When $D/2$ is an odd number the model presents a low 
temperature AF phase and undergoes a mean-field like second order phase transition at 
a finite critical temperature $T_c=1$. The most interesting situation appears when $D/2$ is 
and even number. In this case long range AF order is suppressed at any finite temperature, 
while the ground state is composed by two equal size anti-aligned AF domains; this state
 becomes infinitely degenerated in the thermodynamic limit, but its entropy per spin goes to zero. 
This result shows that the frustration produced by the interplay between long range interactions 
and lattice topology can produce large ground state degeneracies without the generation of 
a ``rough'' free energy landscape.  This result is of importance in spin glasses studies, where 
models defined on an hypercube are used to analize the influence of large
 dimensionalities\cite{Parisi,Stariolo2}. We have shown that this type of lattice can introduce 
effects of large ground state degeneracies which are independent of the presence of disorder, at 
least in the case of long range interactions. Although the contribution of such effects to the 
entropy per spin goes to zero in the thermodynamic limit, its importance could not be negligible for 
finite systems, as in the case of numerical simulation studies. It is worth to stress that the antiferromagnetic model do not have a well defined thermodynamic limit for general dimension of the hypercell $D$. The system exhibit a well defined thermodynamic limit only when the dimension $D$ is restricted to a particular set of values, that is, to be a multiple of four or not. This peculiar behavior illustrates the type of subtleties that may arise when working with long range frustrating interactions in hypercubic cells.

Finally, in the ferromagnetic version of the model we verified a previous conjecture\cite{Cannas2}
 for non-extensive  (ferromagnetic) systems, that is, all the thermodynamic behaviour of the
 model can be exactly described at any temperature by mean field theory if  the thermodynamic
 variables are properly rescaled with the system size.

This work was partially supported by grants from
Consejo Nacional de Investigaciones Cient\'\i ficas y T\'ecnicas CONICET
(Argentina), Agencia C\'ordoba Ciencia (C\'ordoba, Argentina) and  Secretar\'\i
a de Ciencia y
Tecnolog\'\i a de la Universidad Nacional de C\'ordoba (Argentina).  L.F.
acknowledges support from MRC IRC (Oxford, UK). We acknowledge fruitful suggestions of one anonymous referee that helped to improve the quality of the manuscript.

\end{document}